\pdfoutput=1

\documentclass[11pt]{article}

\usepackage[preprint]{acl}

\usepackage{times}
\usepackage{latexsym}
\usepackage[linesnumbered,ruled,vlined]{algorithm2e}

\usepackage[T1]{fontenc}

\usepackage[utf8]{inputenc}

\usepackage{microtype}

\usepackage{inconsolata}

\usepackage{graphicx}

\usepackage[utf8]{inputenc} 
\usepackage[T1]{fontenc}    
\usepackage{hyperref}       
\usepackage{url}            
\usepackage{booktabs}       
\usepackage{amsfonts}       
\usepackage{nicefrac}       
\usepackage{microtype}      
\usepackage{xcolor}         

\usepackage{multirow}
\usepackage{graphicx}
\usepackage{ulem}
\usepackage{caption}
\usepackage{subcaption}
\usepackage{amssymb}
\usepackage{booktabs}
\usepackage{multibib}
\usepackage{colortbl}
\usepackage{makecell}
\usepackage{wrapfig}
\usepackage{color}
\usepackage{float}
\usepackage{inconsolata}
\usepackage{amsmath}

\usepackage{enumitem}
\setlist{leftmargin=1em}

\usepackage[listings]{tcolorbox}
\tcbuselibrary{listings,theorems}
\newtcolorbox{mybox}{colback=white!5!white,colframe=black!75!black, left=.05in, right=.05in}

\definecolor{bluex}{rgb}{0.27, 0.42, 0.81}
\definecolor{purplex}{HTML}{9564bf}
\definecolor{red3}{HTML}{C52A20}
\definecolor{red2}{HTML}{B36A6F}
\definecolor{red1}{HTML}{FFb5b5}
\definecolor{purple}{HTML}{B36A6F}
\definecolor{darkyellow}{HTML}{D5BA82}
\definecolor{blue1}{HTML}{508AB2}
\definecolor{blue2}{HTML}{C4E4E3}
\definecolor{green1}{HTML}{A1D0C7}
\definecolor{green2}{HTML}{BFF6BA}
\definecolor{green3}{HTML}{028100}
\definecolor{teal}{HTML}{508AB2}
\definecolor{purple1}{HTML}{8d3a94}

\newtcbtheorem[]{exmp}{Example}%
{colback=purplex!5, colframe=purplex!80, fonttitle=\bfseries, 
 width=\textwidth, 
 left=0.1in, right=0.1in, bottom=0.05in, top=0.05in}{exmp}

\title{SoRFT: Issue Resolving with Subtask-oriented Reinforced Fine-Tuning}

\author{Zexiong Ma\thanks{\, Work done during the internship at ByteDance.}\hspace{0.4mm} $^{\diamondsuit}$,\, Chao Peng\thanks{\, Corresponding authors.}$^{\clubsuit}$,\, Pengfei Gao\hspace{0.4mm} $^{\clubsuit}$,Xiangxin Meng\hspace{0.4mm} $^{\clubsuit}$,\\\textbf{Yanzhen Zou}$^{\diamondsuit}$,\,  \textbf{Bing Xie}$^{\dagger\diamondsuit}$
\vspace{1mm}\\
  $^{\diamondsuit}$School of Computer Science, Peking University,\,\,
  $^{\clubsuit}$ByteDance\vspace{1mm}\\
  $^{\diamondsuit}$\texttt{mazexiong@stu.pku.edu.cn, \{zouyz, xiebing\}@pku.edu.cn}, \\
  $^{\clubsuit}$\texttt{\{pengchao.x, gaopengfei.se, mengxiangxin.1219\}@bytedance.com}
}

\begin{document}
\maketitle
\begin{abstract}

Mainstream issue-resolving frameworks predominantly rely on commercial models, leading to high costs and privacy concerns. Existing training approaches for issue resolving struggle with poor generalization and fail to fully leverage open-source development resources. We propose \textbf{S}ubtask-\textbf{o}riented \textbf{R}einforced \textbf{F}ine-\textbf{T}uning (\textbf{SoRFT}), a novel training approach to enhance the issue resolving capability of LLMs. We decomposes issue resolving into structured subtasks: file localization, function localization, line localization, and code edit generation. SoRFT consists of two training stages: (1) \textbf{rejection-sampled supervised fine-tuning}, Chain of Thought (CoT) data is filtered using ground-truth before fine-tuning the LLM, and (2) \textbf{rule-based reinforcement learning}, which leverages PPO with ground-truth based rewards. We evaluate the SoRFT-trained model on SWE-Bench Verified and SWE-Bench Lite, achieving state-of-the-art (SOTA) performance among open-source models (e.g., resolve 21.4\% issues on SWE-Bench Verified with SoRFT-Qwen-7B). The experimental results demonstrate that SoRFT significantly enhances issue-resolving performance, improves model generalization, and provides a cost-efficient alternative to commercial models.
\end{abstract}

\section{Introduction}
Large Language Models (LLMs)\cite{gpt4, touvron2023llama} have demonstrated exceptional performance across a wide range of complex real-world tasks \cite{li2022competition, li2024can, wu2023autogen}, particularly excelling in software development \cite{jimenez2024swebench, zhao2024commit0, ma2024compositional}. The current mainstream automated software development systems mainly use commercial models \cite{gpt4, Claude}. However, the API call of commercial models are costly and pose privacy leakage issues, limiting their application in development processes in the industry.

\begin{figure}[t]
\centering
\includegraphics[width=0.99\columnwidth]{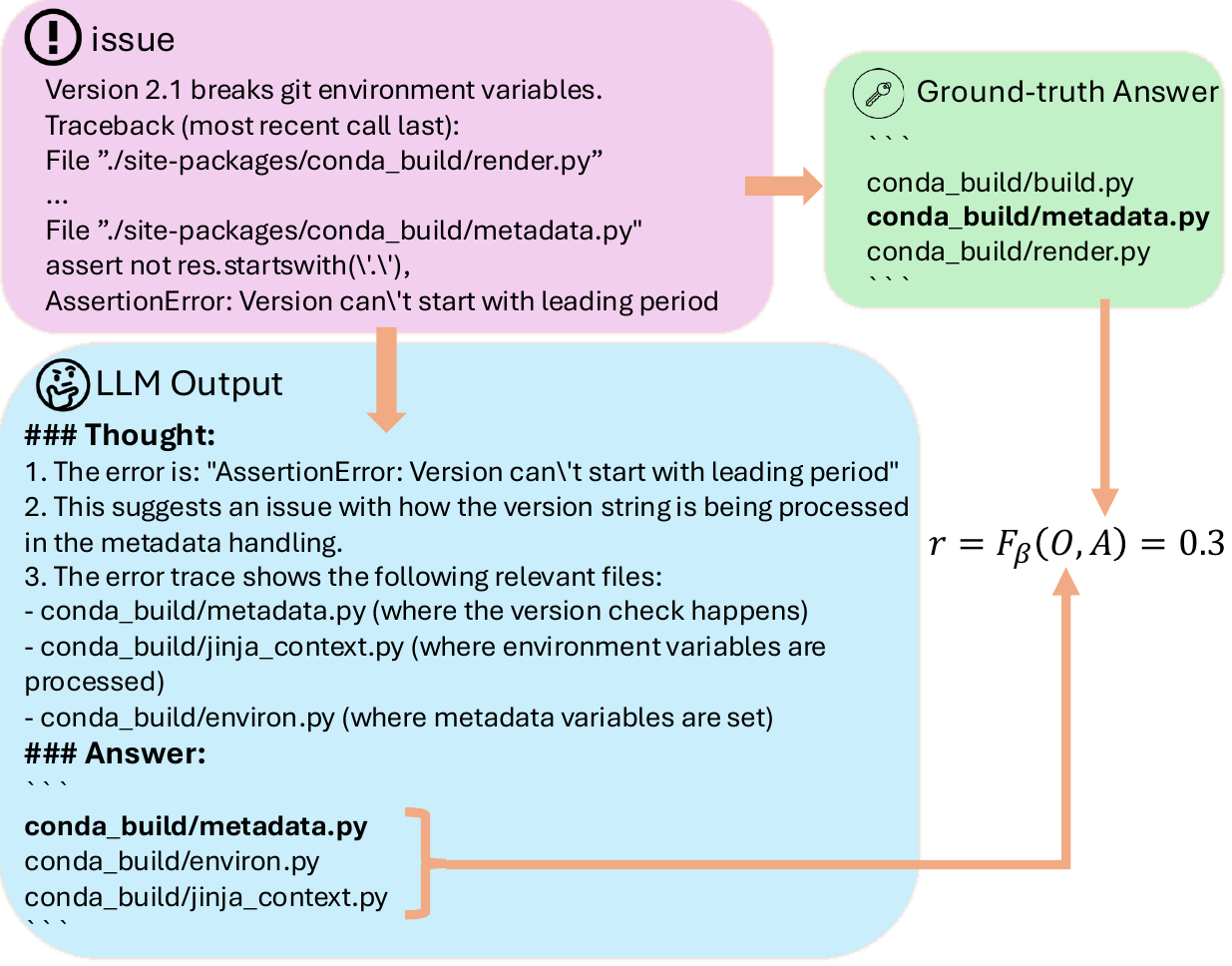}
\vspace{-6mm}
\caption{Rule-based reward example for file localization subtask. LLM generates CoT data for a given issue, the reward for the sampled CoT is then calculated by the $F_\beta$ score based on the extracted answer and the ground-truth answer.}
\label{fig:reward_example}
\end{figure}

Research communities have attempted to fine-tune open-source LLMs \cite{codeqwen1.5, guo2024deepseek, CodeQwen1.5-7B-OpenDevin, touvron2023llama, hui2024qwen2, lozhkov2024starcoder} to improve their performance in Issue Resolving task. Existing Approaches\cite{pan2024training, ma2024repository, xie2025swe, ma2024lingma} utilize LLMs to sample Chain-of-Thought (CoT)\cite{wei2022chain} data, then perform Negative Sample Filtering based on ground-truth data, and fine-tune the models. While these methods improve LLMs' issue-resolving capabilities, relying solely on supervised fine-tuning (SFT) can lead to poor generalization, making models more susceptible to hallucinations and factual errors. 

Recent studies~\cite{luong2024reft, guo2025deepseek, zeng2025simplerl, logic-rl, mu2024rule, team2025kimi} have explored rule-based reinforcement learning to enhance model performance on complex tasks such as mathematics. Rule-based reinforcement learning requires ground-truth for evaluation, but constructing ground-truth for math problems is labor-intensive. In the open-source community, a vast number of resolved issues come with ground-truth patches. This raises a natural question: \textit{Can we leverage these (issue, patch) pairs for rule-based reinforcement learning to improve the issue-resolving capabilities of language models?}

In this paper, we propose \textbf{S}ubtask-\textbf{o}riented \textbf{R}einforced \textbf{F}ine-\textbf{T}uning (\textbf{SoRFT}), fully utilizes both positive and negative sample information to improve model performance in Issue Resolving. Given the complexity of the Issue Resolving task~\cite{jimenez2024swebench}, constructing end-to-end training data is challenging. Inspired by Agentless~\cite{xia2024agentless}, we break down issue resolving into multiple subtasks: file localization, function localization, line localization, and code edit generation—and derive ground-truth answers for each subtask from the pull requests associated with the issues. 
We then perform Reinforced Fine-Tuning~\cite{luong2024reft} for these subtasks. 

SoRFT consists of two training stages: \textbf{rejection-sampled supervised fine-tuning} (SFT) and \textbf{rule-based reinforcement learning} (RL). In the SFT stage, we employ a teacher LLM to generate CoT data for each subtask and filter negative samples based on ground-truth answers. We then perform supervised fine-tuning to help the model grasp the subtask structures, underlying reasoning mechanisms, and output formats essential for issue resolving. In the RL stage, since each subtask has a corresponding ground-truth, we employ rule-based proximal policy optimization (PPO)~\cite{schulman2017proximal} for training. Specifically, we define scoring rules based on ground-truth for each subtask, and update the LLM’s parameters using reward-based optimization. This process further improves the model's issue-resolving performance. SoRFT-trained LLMs achieve state-of-the-art performance on SWE-Bench Verified and SWE-Bench Lite, demonstrating the effectiveness of SoRFT in enhancing issue-resolving capabilities.
In this paper, we make the following contributions:
\begin{itemize}
    \item We introduce Subtask-oriented Reinforced Fine-Tuning (SoRFT), designing rule-based rewards for each issue-resolving subtask and enhancing LLMs' issue-resolving capabilities through reinforced fine-tuning.
    \item We apply SoRFT to open-source models and validate its effectiveness within Agentless framework.
    \item We investigate the impact of different reward rules on PPO training, providing insights for designing more robust reward rules.
\end{itemize}

\section{Background}\label{sec:background}

In this section, we provide a brief introduction to SWE-Bench, issue resolving framework, and the reinforcement learning algorithm.


\subsection{SWE-Bench}

SWE-Bench~\cite{jimenez2024swebench} is a benchmark to evaluate language models' ability to resolve real-world software issues, such as bug reports and feature requests on GitHub. LLM-based programming assistants are given an issue description along with the entire repository and are expected to generate code edits that resolve the issue.
SWE-Bench Lite~\cite{swebench-lite} is a curated subset of SWE-Bench, specifically focus on evaluating functional bug fixes. While SWE-Bench Verified~\cite{swebench-verified} is a human-verified subset addressing quality issues in SWE-Bench, such as vague problem descriptions.
All issue-resolving experiments in this paper are conducted on SWE-Bench Lite and SWE-Bench Verified.

\begin{figure*}[t]
\centering
\includegraphics[width=.93\textwidth]{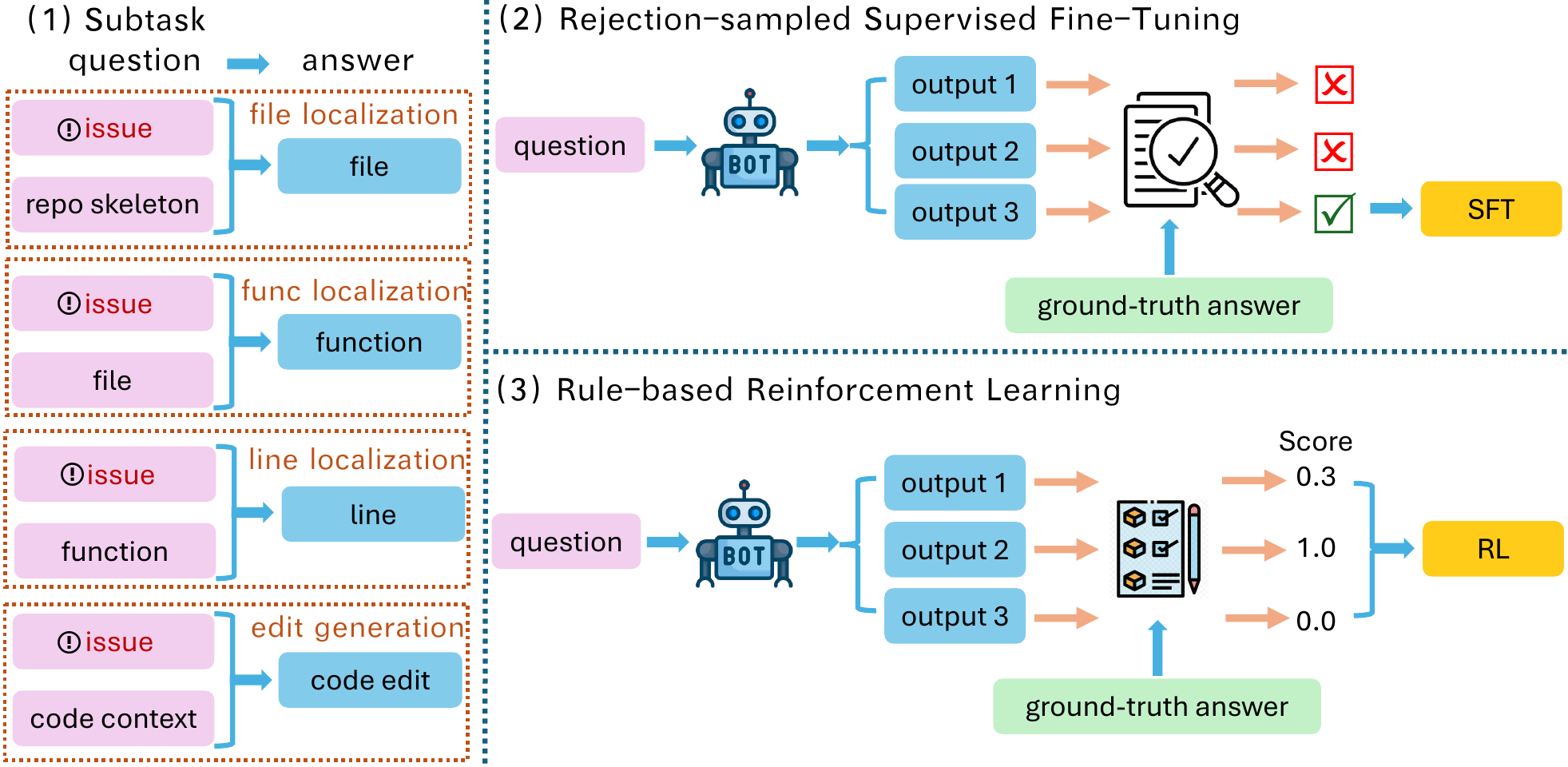}
\caption{SoRFT consists three parts: (1) decompose issue resolving into four subtasks: file localization, function localization, line localization and code edit generation; (2) fine-tune LLMs with rejection-sampled CoT data to enable it follow the task format and reasoning methods for each subtask; (3) employ rule-based reinforcement learning to further enhance the issue resolving ability of LLMs.}
\label{fig:approach}
\end{figure*}

\subsection{Issue Resolving Framework}
Issue resolving frameworks can be broadly divided into two categories: agent-based and pipeline-based. Openhands\cite{wang2024openhands} is a purely react-style~\cite{yao2022react} agent-based framework for software development tasks. \citet{xie2025swe} propose SWE-Fixer, a two-stage pipeline-based system. \citet{ma2024lingma} propose SWE-SynInfer, a hybrid framework that combines pipeline design with agent-based approaches. Additionally,  \citet{xia2024agentless} propose Agentless, a multi-stage pipeline-based framework designed to fully leverage the reasoning capabilities of LLMs for issue resolving. Notably, Agentless is the state-of-the-art (SOTA) pipeline-based framework on the SWE-Bench leaderboard and has been adopted by OpenAI \cite{o1-preview} and DeepSeek \cite{guo2025deepseek} to assess their models' performance in software engineering tasks.
Constructing training data for issue-resolving framework requires sampling CoT data for the framework and filtering out negative samples. Assessing the accuracy of intermediate steps in the Agent framework is challenging, whereas the pipeline-based framework offers a clearer and more effective approach to evaluating reasoning accuracy at each stage. In this paper, we design training sub-tasks based on Agentless and employ Agentless\footnote{We employ Agentless 1.0 in our paper.} as the inference framework in our experiments.

\subsection{Reinforcement Learning}

Reinforcement learning algorithms are widely used in the alignment phase of large language models (LLMs), including proximal policy optimization (PPO)~\cite{schulman2017proximal}, group relative policy optimization (GRPO)~\cite{shao2024deepseekmath}, direct preference optimization (DPO)~\cite{rafailov2023direct}, and Kahneman-Tversky optimization (KTO)~\cite{ethayarajh2023halos, ethayarajh2024kto}. The PPO algorithm calculates the reward using Equation (1) and subsequently updates the parameters of the Policy Model.
\begin{equation}
    r = r_\phi(q, o) - \beta \cdot \text{KL}[\pi_\theta(\cdot|q) \parallel \pi_{ref}(\cdot|q)]
\end{equation}
where $q$ represents the given question, $o$ refers to the predicted output, $r_\phi$ is the reward model, $\pi_\theta$ is the policy model, $\pi_{ref}$ is the reference model (mostly refer to the original policy model $\pi_{\theta_{old}}$). As PPO is capable of maintaining stability and efficiency during policy updates, we employ PPO for RL training in this paper. 

In the standard PPO algorithm~\cite{schulman2017proximal}, a pre-trained reward model is used to score responses. However, recent studies~\cite{guo2025deepseek, team2025kimi} have demonstrated that relying on a reward model can lead to reward hacking, and adopting a more objective scoring approach can effectively mitigate this issue. Since ground-truth can be easily collected for issue resolving tasks, we designed a set of scoring rules to evaluate responses specifically for the issue resolving task. We will present the reward rules in Section~\ref{sec:reward_rule}.

\section{Approach}\label{sec:approach}

As shown in Figure~\ref{fig:approach}, SoRFT contains three parts: (1) issue resolving subtask construction, (2) rejection-sampled supervised fine-tuning, and (3) rule-based reinforcement learning.

\subsection{Issue Resolving Subtasks}

To address the challenge of constructing end-to-end training data for the Issue Resolving task, we create training data for phased subtasks. As shown in Figure~\ref{fig:approach}(1), inspired by the design patterns of advanced issue-resolving frameworks~\cite{xia2024agentless, ma2024lingma}, we decompose issue resolving into four subtasks: file localization, function localization, line localization, and code edit generation.
This structured decomposition enables a more targeted and effective training process for each phase of the task.

\paragraph{File Localization.}

File localization training enhances the LLMs' understanding of the high-level architecture of the repository, enabling them to perform an initial rough localization of relevant files based on the issue description. We utilize the issue and repository structure as inputs, with the full names of the modified files in the pull request (PR) as outputs. To improve the quality of the training data, we excluded non-Python files and test scripts from both input and output. The relationship can be formulated as follows:

\begin{equation} \text{prompt}_{\text{file}}(\mathcal{I}, \mathcal{R}_s) \rightarrow \mathcal{F}_g  \end{equation}
where $\mathcal{F}_g$ represents the golden file, $\mathcal{I}$ represents the issue, $\mathcal{R}_s$ represents the repository skeleton.

\paragraph{Function Localization.}

Function localization training can improve the LLMs' performance on fine-grained localization based on the functional characteristics of the code. In function localization, the issue and file skeleton composed of function names are employed as inputs, while the names of modified functions from the PRs are employed as outputs. This relationship is expressed as:

\begin{equation} \text{prompt}_{\text{function}}(\mathcal{I}, \mathcal{F}_{g,s}) \rightarrow \mathcal{FN}_g \end{equation}
where $\mathcal{FN}_g$ represents the golden function name, $\mathcal{I}$ represents the issue, $\mathcal{F}_{g,s}$ represents the skeleton of the golden file $\mathcal{F}_g$.

\paragraph{Line Localization.}

Line localization training enhances the LLMs' ability to precisely identify the exact lines of code that require modification to resolve the issue. Line localization takes the issue description and function content as inputs and outputs the modified lines from the PR. This can be formulated as:

\begin{equation} \text{prompt}_{\text{line}}(\mathcal{I}, \mathcal{FN}_{g,c}) \rightarrow \mathcal{L}_g \end{equation}
where $\mathcal{L}_g$ represents the golden modified line, $\mathcal{I}$ represents the issue, $\mathcal{FN}_{g,c}$  represents the content of the golden function $\mathcal{FN}_g$. 

\paragraph{Code Edit Generation.}

Training code edit generation can enhance the LLMs' ability to modify code snippets based on the issue. The input for code edit consists of the issue and the localized code snippet, while the output is the code edits of the corresponding PR. Following previous work~\cite{xia2024agentless, yang2024swe}, we employ the \textit{Search/Replace} edit format. The \textit{Search/Replace} format consists of two main parts: 1) Search: the target code snippet that need to modify, and 2) Replace: the new code snippet after editing. This relationship can be formulated as:

\begin{equation} \text{prompt}_{\text{edit}}(\mathcal{I}, \mathcal{C}_{g}) \rightarrow  \mathcal{E}_g\end{equation}
where $\mathcal{E}_g$ represents the golden code edit, $\mathcal{I}$ represents the issue, $\mathcal{C}_g$ represents the golden code context.

All subtasks are constructed based on resolved issues from open-source projects, and the ground-truth answers are extracted from the corresponding pull requests. The prompts for the subtasks are provided in Appendix~\ref{sec:subtask_prompt}. 

\subsection{Rejection-sampled Supervised Fine-Tuning}

We fine-tune the LLM using Rejection-sampled CoT data to enhance its understanding of the task format and reasoning process for each subtask. As shown in Figure~\ref{fig:approach}(2), we sample CoT data using the LLM and then filter the CoT data based on the ground-truth answer. Specifically, for the three localization subtasks, we filter out samples that have no overlap with the ground-truth file, function or line. For the code edit generation subtask, we filter out samples that have no overlap with the lines modified by the ground-truth edits. Finally, we integrate CoT data from all subtasks to fine-tune the LLM, enabling it to comprehend both the task format and its underlying reasoning mechanisms.

\begin{algorithm} [t]
\footnotesize
\caption{Rule-based reward}
\label{algo:reward}
\KwIn{Subtask type $s$, question prompt $q$, ground-truth answer $A$, LLM output $o$}
\KwOut{Reward score $r$}
\If{$s$ == \textit{localization}}{ \label{algo:location_begin}
    $O_{loc} \leftarrow extract\_locations(o)$;\\
    
    $Q \leftarrow extract\_locations(q)$; \\
    \If{$|O_{loc}| == 0$ or $|O_{loc}-Q| > 0$}{
        $r = 0.0$ 
    }\Else{
        $r = F_\beta(O_{loc},A)$ 
    }
    
}\label{algo:location_end}
\If{$s$ == \textit{edit}}{ \label{algo:edit_begin}
    $O_{search} \leftarrow extract\_search\_blocks(o)$;\\
    $O_{edit} \leftarrow extract\_edits(o)$;\\
    $Q \leftarrow extract\_code(q)$; \\
    \If{$|O_{edit}| == 0$ or $|O_{search}-Q| > 0$}{
        $r = 0.0$
    }\Else{
        $r = F_\beta(O_{edit},A)$
    } 
    
} \label{algo:edit_end}
\Return{$r$}
\end{algorithm}

\subsection{Ruled-based Reinforcement Learning}\label{sec:reward_rule}

We further enhance the reasoning ability and generalization of LLMs on issue-resolving through Rule-based Reinforcement Learning. As shown in Algorithm~\ref{algo:reward}, we utilize the ground-truth answer to calculate the rule-based reward for each subtask. For the localization subtask (line~\ref{algo:location_begin}-\ref{algo:location_end}), we first extract the localization result $O_{loc}$ from the response. If the localization result is empty or contains a target not present in the problem description, the reward is set to 0; otherwise, the reward is calculated as the $F_\beta$  score between $O_{loc}$ and the ground-truth answer $A$. For the code editing subtask (line~\ref{algo:edit_begin}-\ref{algo:edit_end}), we first extract the modification target $O_{search}$ and the modification code $O_{edit}$ from the response. If the modification code is empty or the modification target does not appear in the problem description, the reward is 0; otherwise, the reward is calculated as the $F_\beta$  score between $O_{edit}$ and the ground-truth answer $A$. During reinforcement learning period, we replace the reward model in PPO with above rule-based reward. This approach effectively mitigates the risk of reward hacking, ensuring that the model's learning process is guided by precise and reliable feedback.
\begin{equation} 
F_\beta = (1 + \beta^2) \cdot \frac{\text{Precision} \cdot \text{Recall}}{(\beta^2 \cdot \text{Precision}) + \text{Recall}}, \\
\end{equation}
\begin{equation}
\text{Precision} = \frac{|O \cap A|}{|O|}, \\
\text{Recall} = \frac{|O \cap A|}{|A|}
\end{equation}
where $O$ represents the outputs generated by the LLMs, $A$ denotes the ground-truth answers for the subtask.  $\beta$ is a hyperparameter that balances the impact of precision and recall on final score. Since recall has a greater influence on the final outcome across different subtasks, $\beta$ should be a value greater than 1. In our experiments, we set $\beta=3$ to prioritize recall while maintaining a reasonable trade-off with precision.

\section{Experiments}

In this section, we will introduce our evaluation benchmark, metrics, models, baselines and implementation details.

\subsection{Benchmark}
We conduct experiments on two issue resolving benchmarks: SWE-Bench Verified and SWE-Bench Lite.

\textbf{SWE-Bench Verified}~\cite{swebench-verified} is a manually verified subset of SWE-bench~\cite{jimenez2024swebench} with 500 instances. Each instance is a issue associated with test cases that can be executed in a Docker\footnote{https://www.docker.com/} environment. Issue resolving frameworks~\cite{xia2024agentless, xie2025swe, wang2024openhands, ma2024lingma} are asked to understand the issue and the repository, generate patches and pass all test cases, providing a reliable evaluation of their issue resolving capabilities.

\textbf{SWE-Bench Lite}~\cite{swebench-lite} is the subset of SWE-Bench containing 300 instances and focuses on evaluating functional bug fixes.

\subsection{Metrics}
To evaluate the performance of issue resolving frameworks with SoRFT-trained LLMs, we apply two metrics: \textbf{\%Resolved}, \textbf{\%Applied}. 

\textbf{\%Resolved} is the proportion of samples in which applying the generated code edit successfully passed all test cases.

\textbf{\%Applied} is the proportion of samples that issue resolving frameworks successfully generate valid code edits that could be applied to the repositories. This metric evaluates the framework's capability to produce practical and executable solutions.

\begin{table*}[tbp]
\caption{The \%Resolved performance of various models on SWE-Bench Verified and SWE-Bench Lite. Given that all fine-tuning approaches are inherently framework-specific, we compare SoRFT-Qwen with previous fine-tuned models within corresponding frameworks.}
\label{tab:RQ1}
\centering
\scalebox{0.8}{
\begin{tabular}{llccc}
\toprule
\textbf{Model}         & \textbf{Framework}                & \textbf{Type} & \textbf{Verified} & \textbf{Lite} \\
\midrule
\multicolumn{5}{c}{\cellcolor{gray!11}{\textbf{Proprietary Models}}} \\
\midrule
Claude-3.5-Sonnet~\cite{Claude}      &  Openhands          & Agent & 53.0           & 41.7             \\
Claude-3.5-Sonnet~\cite{Claude}        & Agentless       & Pipeline & 50.8           & 40.7            \\
GPT-4o~\cite{gpt4o}        & SWE-SynInfer        & Pipeline + Agent & 31.8          & 20.7            \\
\midrule
\multicolumn{5}{c}{\cellcolor{gray!11}{\textbf{7 - 14B Open-source Models}}} \\
\midrule
SWE-Gym-Qwen-7B~\cite{pan2024training}        &  Openhands          & Agent & 10.6           & 10.0             \\
SWE-Gym-Qwen-14B~\cite{pan2024training}        & Openhands          & Agent & 16.4            & 12.7             \\
Lingma-SWE-GPT-7B~\citep{ma2024lingma}        & SWE-SynInfer            & Pipeline + Agent & 18.2           & 12.0      \\
\cellcolor{blue!11}{\textbf{SoRFT-Qwen-7B (Ours)}} & \cellcolor{blue!11}{Agentless}     & \cellcolor{blue!11}{Pipeline} & \cellcolor{blue!11}{\textbf{21.4}} & \cellcolor{blue!11}{\textbf{14.0}} \\

\midrule
\multicolumn{5}{c}{\cellcolor{gray!11}{\textbf{32 - 72B Open-source Models}}} \\
\midrule

Lingma-SWE-GPT-72B~\citep{ma2024lingma}       & SWE-SynInfer            & Pipeline + Agent & 30.2           & 22.0       \\
SWE-Fixer-Qwen-72B~\citep{xie2025swe}                  & SWE-Fixer   &  Pipeline & 30.2          & 23.3             \\
SWE-Gym-Qwen-32B~\cite{pan2024training}        & Openhands         &  Agent & 20.6            & 15.3             \\
\cellcolor{blue!11}{\textbf{SoRFT-Qwen-32B (Ours)}} & \cellcolor{blue!11}{Agentless}     &  \cellcolor{blue!11}{Pipeline} & \cellcolor{blue!11}{\textbf{30.8}} & \cellcolor{blue!11}{\textbf{24.0}} \\
\bottomrule

\end{tabular}
}
\end{table*}

\subsection{Framework and Model}
We apply Agentless\cite{xia2024agentless} as our issue resolving framework and Qwen2.5-Coder~\cite{hui2024qwen2} series as our base model. Agentless is an advanced open-source issue-resolving framework, used by OpenAI~\cite{o1-preview} and DeepSeek~\cite{guo2025deepseek} to evaluate model performance on software engineering tasks. For base model, we employ Qwen2.5-Coder-7B-Instruct and Qwen2.5-Coder-32B-Instruct\cite{hui2024qwen2}, which is the SOTA open-source coder instruct models~\cite{touvron2023llama, guo2024deepseek}.

\subsection{Baselines}
Since issue resolving tasks necessitate the use of agent-based or pipeline-based frameworks, existing fine-tuning approaches are typically designed and optimized for specific frameworks. In this work, we evaluate our method against three baselines: (1) OpenHands with SWE-Gym-Qwen, (2) SWE-Fixer with SWE-Fixer-Qwen, and (3) SWE-SynInfer with Lingma-SWE-GPT.

\paragraph{Openhands with SWE-Gym-Qwen\cite{pan2024training}.} Openhands\cite{wang2024executable} is a purely React-style\cite{yao2022react} agent-based framework, equipped with tools such as file viewing and bash command execution. The framework enables the model to autonomously invoke these tools in a React-like manner, iteratively reasoning and acting to resolve issues. They collected trajectories of Openhands invoking GPT-4o\cite{gpt4o} and Claude-3.5-Sonnet\cite{Claude} for issue resolving tasks, filtered out the failed trajectories, and then fine-tuned the Qwen2.5-Coder model to serve as the SWE-Gym-Qwen model.

\paragraph{SWE-Fixer with SWE-Fixer-Qwen\cite{xie2025swe}.} SWE-Fixer is a two-stage pipeline-based framework. First, it uses a retriever that combines BM-25 and LLMs to locate the files to be modified, and then uses LLMs to generate code edits to the files. They utilized GPT-4o\cite{gpt4o} to collect CoT training data, and fine-tuned the Qwen2.5-Coder model to serve as the SWE-Fixer-Qwen model.

\paragraph{SWE-SynInfer with Lingma-SWE-GPT\cite{ma2024lingma}.} SWE-SynInfer is a hybrid framework that combines pipeline design with agent-based capabilities. In the first stage, the model sequentially analyzes the repository structure, file skeletons, and code snippets to generate a detailed modification plan. Then, it provides tools such as file viewing, allowing the model to invoke these tools in a react manner and generate the final code edit. Similar to Openhands\cite{wang2024executable}, they collected trajectories of SWE-SynInfer invoking GPT-4o\cite{gpt4o} for issue resolving tasks, filtered out the failed trajectories, and then fine-tuned the Qwen2.5-Coder model to serve as the Lingma-SWE-GPT model.

\subsection{Implementation Details}
In this subsection, we will introduce the data construction details, fine-tuning details, and reinforcement learning details.

\paragraph{Data Construction.} To construct our training dataset, we curate a collection of high-quality open-source Python projects from \textit{seart-ghs}~\footnote{https://seart-ghs.si.usi.ch/} by applying a set of stringent criteria. Specifically, we select repositories that satisfy the following conditions: (1) at least 1,000 issues, (2) at least 1,000 pull requests (PRs), (3) a minimum of 100 stars, (4) inclusion of an appropriate license, (5) exclusion of forked repositories, and (6) absence from the SWE-Bench test dataset to prevent data contamination. Through this rigorous selection process, we identify a final set of 660 repositories.
From this pool, we further select 100 repositories to generate the SFT CoT data. We extract 30,000 (issue, PR) pairs from these repositories and formulate corresponding subtasks. We sample CoT data using Claude-3.5-Sonnet~\cite{Claude} and filter out instances where the final answer does not align with the subtask's ground truth, retaining 60k training samples. We also crawl the (issue, PR) data from the remaining repositories and construct the corresponding subtasks. During the reinforcement learning phase, we randomly select 30k samples for training.

\begin{figure*}
    \centering
    \subfloat[The response length curve during RL training with hit score.]{
    \label{fig:hit_score} 
    \includegraphics[width=0.27\textwidth]{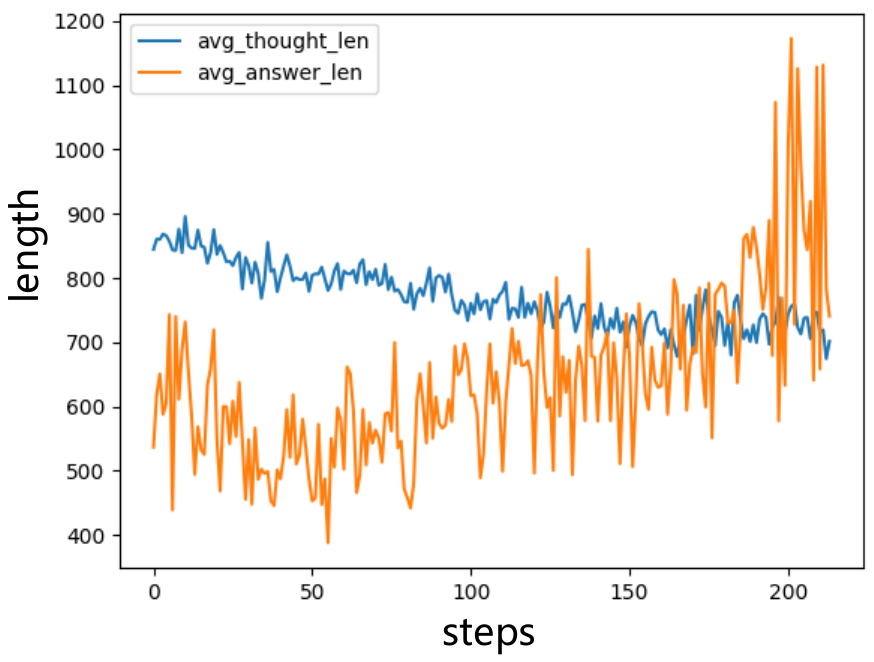}}\quad
    \subfloat[The response length curve during RL training with $F_\beta$ Score.]{
    \label{fig:f3_score} 
    \includegraphics[width=0.27\textwidth]{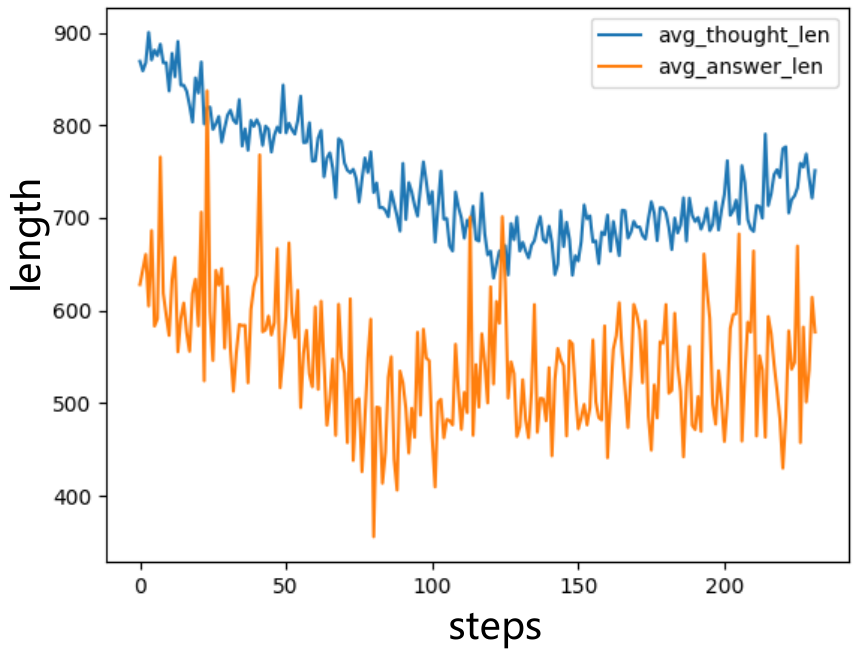}}\quad
    \subfloat[Performance comparison of different rule-based reward on SWE-bench Verified.]{
    \label{fig:reward_ablation} 
    \includegraphics[width=0.37\textwidth]{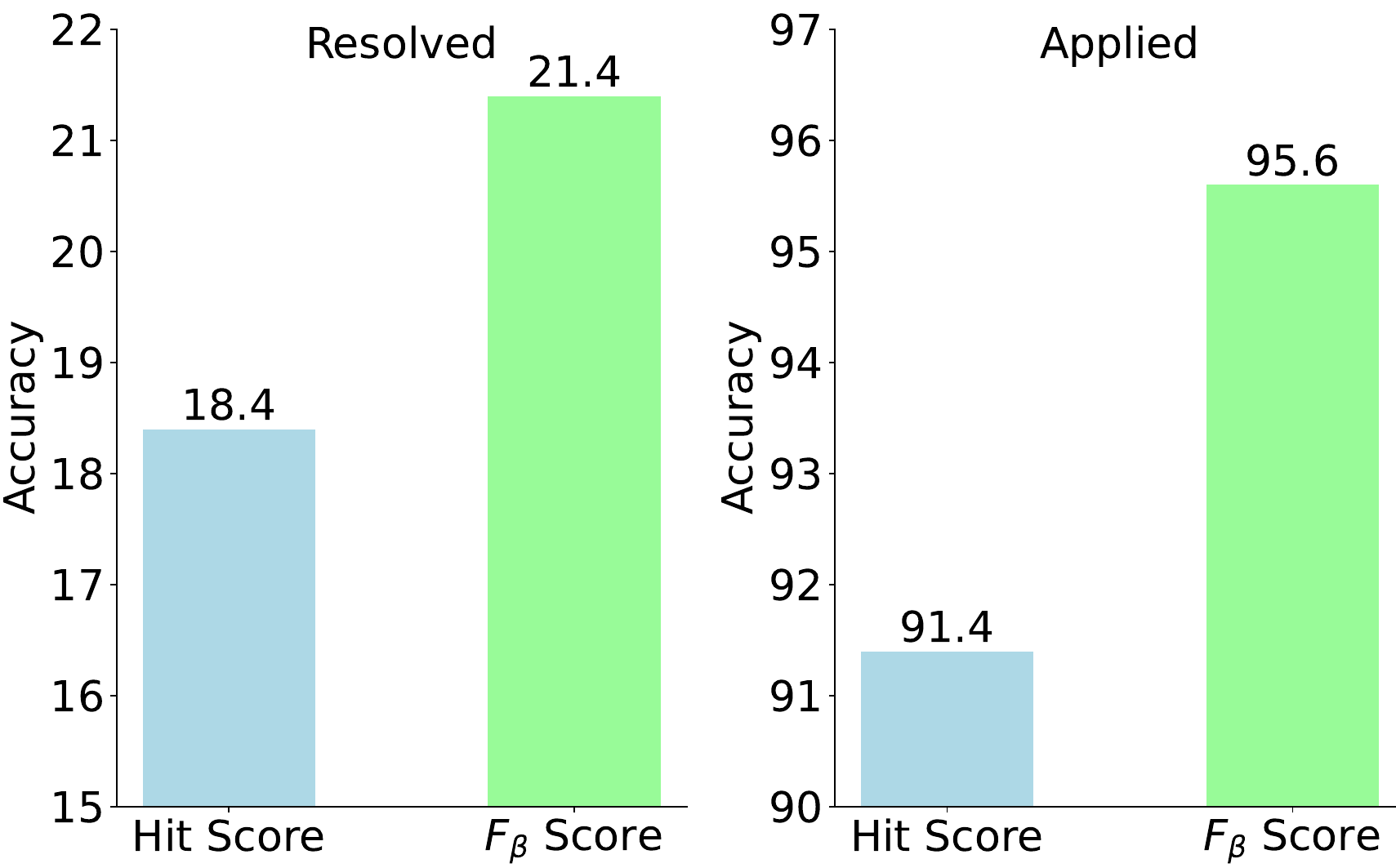}}\quad
    
    \caption{Comparison of rule-based reward strategy: hit score v.s. $F_\beta$ score.}
    \label{fig:APIRec_ablation}
\end{figure*}

\paragraph{Fine-tuning.} We employ FastChat~\cite{zheng2023judging} framework  with full sharding strategy and CPU offload strategy implemented by Pytorch FSDP~\footnote{https://pytorch.org/docs/stable/fsdp.html} for our 7B model fine-tuning, and employ DeepSpeed~\cite{rasley2020deepspeed} for our 32B model fine-tuning. The training process is conducted on 4x8 96G H20 GPUs. We also utilize flash-attention-2~\cite{dao2023flashattention2} to reduce memory overhead and speed up the training process. We set the global batch size to 128 and train for 2 epochs. We apply cosine learning rate decay with a maximum learning rate of 1e-5 and 3\% warm-up steps.

\paragraph{Reinforcement Learning.} We employ OpenRLHF~\cite{hu2024openrlhf} framework for our PPO~\cite{schulman2017proximal} implementation. The training process is conducted on 4x8 96G H20 GPUs. We also utilize ray~\cite{moritz2018ray}, DeepSpeed~\cite{rasley2020deepspeed}, flash-attention-2~\cite{dao2023flashattention2} and vllm~\cite{kwon2023efficient} to reduce GPU memory overhead and speed up the training process. We set temperature to 1.0 to sample completions for each prompt.

\section{Results and Analysis}

\paragraph{SoRFT achieves SOTA performance among open-source LLMs.} Table ~\ref{tab:RQ1} presents the performance of SoRFT and fine-tuned open-source LLMs on the issue-resolving task on SWE-bench Verified and SWE-bench Lite. We categorize the open-source models into two groups based on their parameter sizes: (1) 7-14B open-source LLMs, and (2) 32-72B open-source LLMs. The LLMs trained with SoRFT achieved state-of-the-art (SOTA) performance among models of the same parameter size and even slightly outperforms some larger models. On SWE-bench Verified, SoRFT-Qwen-7B outperforms SWE-Gym-Qwen-32B (21.4 vs. 20.6). SoRFT-Qwen-32B even outperforms Lingma-SWE-GPT-72B (30.8 vs. 30.2), despite the latter having significantly more parameters.

While OpenHands achieves optimal performance with proprietary models, the SWE-Gym model, specifically fine-tuned for OpenHands, underperforms compared to others. This discrepancy may arise from the challenges of constructing supervision signals for intermediate steps in agent framework, whereas a pipeline framework can establish supervision signals for different stages. This allows for finer-grained filtering of CoT data and more precise reward calculation.

\begin{table}
    \centering
    \caption{Performance comparison of model with different training strategy on SWE-bench Verified.}
    \label{tab:RQ2}
    \resizebox{.99\linewidth}{!}{
        \begin{tabular}{lcc}
        \hline
        Model & \textbf{\%Resolved} & \textbf{\%Applied} \\
        \hline
        Qwen2.5-Coder-7B-Instruct & 7.6 & 55.6 \\
        \quad + SFT & 18.0 &	85.2 \\
        \quad + SFT + RL (Our SoRFT-Qwen-7B) & \textbf{21.4} &	\textbf{95.6} \\
        \hline
        Qwen2.5-Coder-32B-Instruct & 25.6 & 84.4 \\
        \quad + SFT & 28.8 &	90.6 \\
        \quad + SFT + RL (Our SoRFT-Qwen-32B) & \textbf{30.8} &	\textbf{95.8} \\
        \hline
        
        \end{tabular}
    }
\end{table}

\paragraph{SoRFT achieves higher accuracy than solely supervised fine-tuning.} On SWE-Bench Verified, we evaluate the performance of agentless framework with different models. As demonstrated in Table~\ref{tab:RQ2}, full SoRFT training consistently outperforms SFT alone for both 7B and 32B models. Both \%Resolved and \%Applied metrics indicate that SoRFT enhances the model’s ability to resolve issues.

\paragraph{The robustness of reward rules is crucial for reinforcement learning.} We conduct an ablation study on the reward rule in our algorithm by replacing the $F_\beta$ score with a simpler hit score. Specifically, if the response contains at least one element from the ground-truth answer, the reward score is set to 1.0; otherwise, it is set to 0.0. As shown in Figure~\ref{fig:hit_score}, using the hit score leads to reward hacking, where the model tends to generate fewer thoughts and more answers to increase the likelihood of including a ground-truth element. In contrast, Figure~\ref{fig:f3_score} demonstrates that using the $F_\beta$  score reduces the generation of redundant answers and stabilizes the answer length. The performance gap in Figure~\ref{fig:reward_ablation} further indicates that a robust reward rule is crucial for SoRFT training.
Notably, for the thought length in Figure~\ref{fig:f3_score}, we observe a trend consistent with recent studies~\cite{zeng2025simplerl}: it initially decreases and then increases during PPO training. In the early stages, the model eliminates unnecessary reasoning to streamline its output. As training stabilizes, it gradually increases the depth and complexity of thought process.

\paragraph{SoRFT also enhances performance on general code tasks.} We also conduct evaluations on two general code tasks: LiveCodeBench~\cite{jain2024livecodebench} and RepoQA~\cite{liu2024repoqa}. LiveCodeBench focuses on self-contained code generation scenarios, while RepoQA evaluates the ability of LLMs to extract information from long-context code content. The results in Table~\ref{tab:RQ4} indicate that SoRFT also has the potential to enhance code-related tasks beyond issue resolving. There is a large amount of development process data in the open-source community, which remains untapped in current LLM training process. Approaches like SoRFT have the potential to utilize these data to further improving the capabilities of code LLMs.

\begin{table}
    \centering
    \caption{Performance comparison on LiveCodeBench and RepoQA.}
    \label{tab:RQ4}
    \resizebox{.99\linewidth}{!}{
        \begin{tabular}{l|cc}
        \hline
        Model & LiveCodeBench & RepoQA \\
        \hline
        Qwen2.5-Coder-7B-Instruct & 34.18 & 85.0 \\
        SoRFT-Qwen-7B & \textbf{34.64} &	\textbf{90.0} \\
        \hline
        
        \end{tabular}
    }
\end{table}

\section{Related Work}

\paragraph{LLM Training for Issue Resolving.} To enhance the issue resolving capabilities of open-source LLMs, several research works~\citep{ma2024lingma, xie2025swe, ma2024repository, pan2024training} have attempted to use software development resources from the open-source community to construct training data and fine-tune open-source LLMs.
~\citet{pan2024training} crawled open-source repositories and utilized closed-source models (e.g., GPT-4o~\cite{gpt4o} and Claude-3.5-Sonnet~\cite{Claude}) to generate Openhands~\cite{wang2024executable, wang2024openhands} Agent trajectories, and filtered them through unit tests. Then they used the trajectories to fine-tune the Qwen~\cite{hui2024qwen2} model, enabling it to serve as the base model for Openhands.
~\citet{ma2024lingma} used GPT-4o to generate Agent trajectories on open-source repository issues, and fine-tuned an open-source model with the filtered trajectories.
~\citet{pan2024training} generated CoT data for and edit generation tasks using GPT-4o, and fine-tuned an open-source model to apply it to SWE-Fixer RAG pipeline.
All the above work used SFT to fine-tune models. To the best of our knowledge, we are the first work to leverage reinforced fine-tuning~\cite{luong2024reft} to enhance the issue-resolving capabilities of LLMs.

\paragraph{Reinforcement Learning with Rule-based Reward.}
Since OpenAI released o1~\cite{o1-preview} model, many efforts have attempted to enhance LLMs' long-form reasoning capabilities through rule-based reinforcement learning. DeepSeek's R1~\cite{guo2025deepseek} model with rule-based GRPO~\cite{shao2024deepseekmath} further demonstrates the potential of rule-based rewards. ~\citet{team2025kimi} released Kimi-k1.5, also trained with rule-based reinforcement learning. The research community~\cite{zeng2025simplerl,logic-rl} has also been working on replicating rule-based reinforcement learning process. ~\citet{tinyzero} trained a 3B model with PPO~\cite{schulman2017proximal} on the Countdown task and observed "\textit{Aha moment}"~\cite{guo2025deepseek} phenomenon, aligns closely with the behavior of R1. ~\citet{zeng2025simplerl} trained a 7B model using PPO on Math task and observed that response length initially decreased and then increased (similar as the tendency in Figure~\ref{fig:f3_score}). Previous work mainly focused on mathematical tasks, where rewards can be straightforwardly computed based on ground truth. In this paper, we improve the performance of open-source models in issue-resolving framework through subtask-oriented rule-based reinforcement learning.

\section{Conclusion}\label{sec:conclusion}
In this paper, we propose SoRFT, a subtask-oriented reinforced fine-tuning approach that enhances LLMs' issue-resolving capabilities. By leveraging rule-based reinforcement learning, SoRFT improves performance while ensuring better generalization. Our results demonstrate its effectiveness as a cost-efficient alternative to commercial models.

\section{Limitations} \label{sec:limitation}
\paragraph{The False Negatives of Rule-based Rewards.} 
The correct resolution to an issue is often not unique. Relying solely on rule-based rewards by comparing the LLM's response to a single ground truth may incorrectly classify valid solutions as failures. To address this limitation, future work could incorporate unit test execution results as a more objective and fair measure of code edit quality. 

\paragraph{Experiments were conducted only in the Python repositories.} Due to the lack of a multilingual SWE-Bench test set, our experiments were limited to Python repositories. However, since SoRFT is a language-agnostic framework, we believe it has the potential to improve the issue-resolving capabilities of LLMs in other languages.

\bibliography{reference}

\appendix
\clearpage
This is the Appendix of the paper: \textit{SoRFT: Issue Resolving with Subtask-oriented Reinforced Fine-Tuning}.

\section{Evaluation Details}
All evaluation experiments are conducted on 4 96G H20 GPUs with vllm~\cite{kwon2023efficient} framework. We employ the official evaluation scripts~\footnote{https://github.com/swe-bench/SWE-bench} from SWE-Bench repository for SWE-Bench Verified and SWE-Bench Lite.  On LiveCodeBench~\cite{jain2024livecodebench}, we evaluate on the code\_generation\_lite \textit{releave\_v4} version, which contains 713 problems released between May 2023 and Sep 2024. On RepoQA~\cite{liu2024repoqa}, we evaluate on Python repositories with \textit{similarity threshold=0.9}.

\begin{figure}[h]
\centering
\includegraphics[width=0.9\columnwidth]{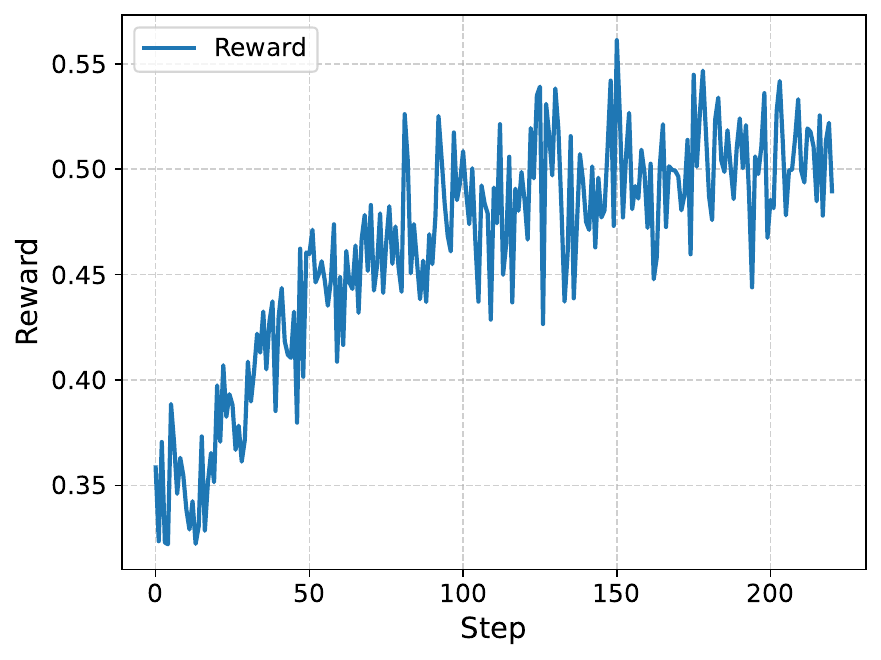}
\caption{Reward over PPO training steps.}
\label{fig:reward_curve}
\end{figure}

\section{PPO Training Details}
To support rule-based PPO training, we implement a customized reward function to replace the original reward model scoring function in the OpenRLHF~\cite{hu2024openrlhf} framework. We set train batch size to 64, the Adam learning rate for the actor model to 5e-7, and the Adam learning rate for the critic model to 9e-6. As shown in Figure~\ref{fig:reward_curve}, using our designed reward rules, the reward steadily increases as the training steps progress.

\section{Additional Reasoning Example}
We observed that the model tends to engage in additional reasoning after SoRFT training on 7B model, which bears some similarity to the "aha moment" observed by DeepSeek during the R1 training process~\cite{guo2025deepseek}. In Example~\ref{exmp:additional_thought}, before concluding its reasoning, the SoRFT-Qwen-7B actively explores whether any potential answers have been overlooked and successfully identifies the ground-truth function \textit{sql\_flush}. 

\section{Subtask Prompt}\label{sec:subtask_prompt}
We designed data generation prompts for different subtasks based on the prompts used at corresponding stages of Agentless~\cite{xia2024agentless}. Each subtask's prompt consists of four parts: subtask description, issue, context, and output format instructions. LLMs are required to analyze the issue description, select content from the context that aligns with the subtask description, and provide the final answer according to the specified output format. 
Example~\ref{exmp:file_localization}-\ref{exmp:code_edit} is the prompt utlized to generate CoT data for each subtask.

\begin{table}
    \centering
    \caption{Performance comparison on issue resolving subtasks.}
    \label{tab:hit_result}
    \resizebox{\linewidth}{!}{
        \begin{tabular}{lccc}
        \hline
        Model & \textbf{\%File Hit} & \textbf{\%Func Hit} & \textbf{\%Line Hit}\\
        \hline
        Qwen2.5-Coder-7B-Instruct & 59.8 & 51.2 & 17.2 \\
        SoRFT-Qwen-7B & \textbf{77.8} &	\textbf{66.4} & \textbf{23.6} \\
        \hline
        
        \end{tabular}
    }
\end{table}

\section{Effectiveness of SoRFT on issue resolving subtasks} 
We conducted a detailed evaluation of the impact of SoRFT on various subtasks of issue resolving using the SWE-Bench Verified dataset. Specifically, we measured the model's hit rates on three subtasks: file localization, function localization, and line localization. The results presented in Table~\ref{tab:hit_result} demonstrate that SoRFT training enhances the model's performance across these subtasks.

        

\begin{figure}[h]
\centering
\includegraphics[width=0.9\columnwidth]{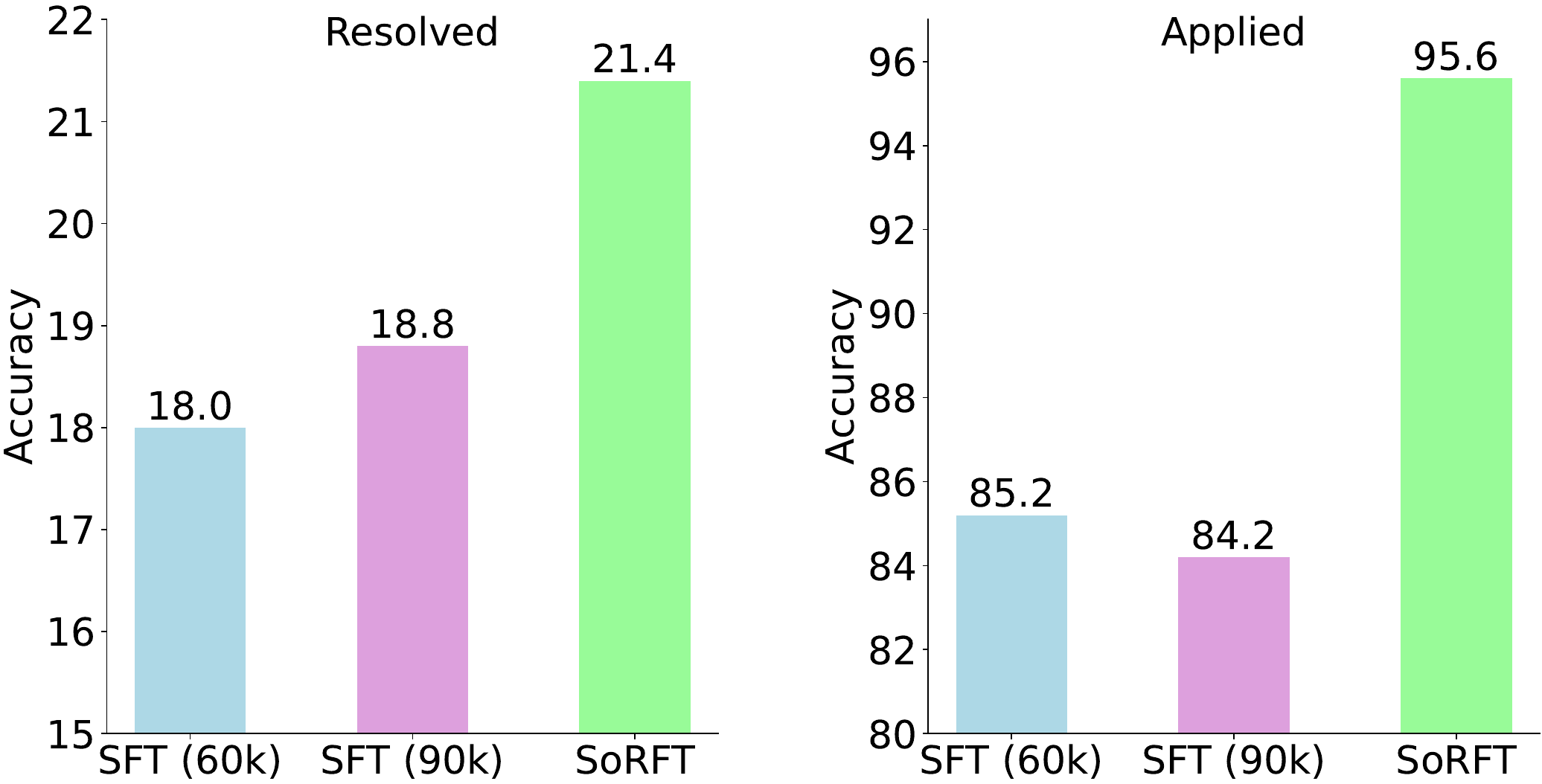}
\caption{Performance of models trained with different training strategies.}
\label{fig:data_ablation}
\end{figure}

\section{Training Data Ablation}
Since the SoRFT utilized more data than SFT in the experiments of Table~\ref{tab:RQ2}, to more fairly evaluate the effectiveness of SoRFT, we conducted an ablation on the amount of training data. As illustrated in Figure~\ref{fig:data_ablation}, even when SFT was trained on the same amount of data as SoRFT (90k samples), its performance still fell short of SoRFT's results. Increasing the training data for SFT from 60k to 90k samples slightly improved the \%resolved but decreased the \%Applied, indicating that excessive SFT might impair the model's generalization ability. In contrast, SoRFT was able to more robustly enhance the model's issue resolving capability.

\begin{figure*}[h]
\begin{exmp}{Additional Reasoning Example}{additional_thought}
\textbf{Input:}\\
Identify all locations that need inspection or editing to fix the problem. \\
\textit{Issue:} Simplify signature of DatabaseOperations.execute\_sql\_flush()\\
...\\
\textbf{Output:}\\
\textit{\#\#\# Thought:} \\
The issue is about simplifying the execute\_sql\_flush method's signature from (self, using, sql\_list) to just (self, sql\_list). \\
...\\
\textbf{Additionally, there's a sql\_flush() method (line 401)} that's related to this functionality and might need to be updated as well.\\
...\\
\textit{\#\#\# Answer:} \\
django/db/backends/base/operations.py \\
class: BaseDatabaseOperations \\
function: sql\_flush \\
...
\end{exmp}
\end{figure*}

\begin{figure*}[h]
    \begin{minipage}{\textwidth} 
        \begin{exmp}{File Localization Prompt}{file_localization}
Please look through the following GitHub problem description and Repository structure and provide a list of files that one would need to edit to fix the problem.\\

\textit{\#\#\# GitHub Problem Description:}\\
\{issue\}

\textit{\#\#\# Repository Structure:}\\
\{repository\_structure\}

Please think step by step, provide the full path and return at most 5 files.\\
The returned files should be separated by new lines ordered by most to least important.\\
For example:\\
\verb|`|\verb|`|\verb|`|\\
file1.py\\
file2.py\\
\verb|`|\verb|`|\verb|`|\\

Your reasoning should start with "\#\#\# Thought:", and your answer should start with "\#\#\# Answer:".
        \end{exmp}
    \end{minipage}
\end{figure*}

\begin{figure*}[h]
    \begin{minipage}{\textwidth} 
        \begin{exmp}{Function Localization Prompt}{func_localization}
Please look through the following GitHub Problem Description and the Skeleton of Relevant Files.\\
Identify all locations that need inspection or editing to fix the problem, including directly related areas as well as any potentially related global variables, functions, and classes.\\
For each location you provide, either give the name of the class, the name of a method in a class, the name of a function, or the name of a global variable.\\

\textit{\#\#\# GitHub Problem Description:}\\
\{issue\}

\textit{\#\#\# Skeleton of Relevant Files:}\\
\{file\_skeleton\}

Please provide the complete set of locations as either a class name, a function name, or a variable name.\\
Note that if you include a class, you do not need to list its specific methods.\\
You can include either the entire class or don't include the class name and instead include specific methods in the class.\\
For example:\\
\verb|`|\verb|`|\verb|`|\\
full\_path1/file1.py\\
function: my\_function\_1\\
class: MyClass1\\
function: MyClass2.my\_method\\
\verb|`|\verb|`|\verb|`|\\

Please think step by step before returning the locations.\\
Your reasoning should start with "\#\#\# Thought:", and your answer should start with "\#\#\# Answer:".
        \end{exmp}
    \end{minipage}
\end{figure*}

\begin{figure*}[h]
    \begin{minipage}{\textwidth} 
        \begin{exmp}{Line Localization Prompt}{line_localization}
Please review the following GitHub problem description and relevant files, and provide a set of locations that need to be edited to fix the issue.\\
The locations can be specified as class names, function or method names, or exact line numbers that require modification.\\

\textit{\#\#\# GitHub Problem Description:}\\
\{issue\}

\textit{\#\#\# File Contents:}\\
\{file\_contents\}

Please provide the class name, function or method name, or the exact line numbers that need to be edited.\\
For example:\\
\verb|`|\verb|`|\verb|`|\\
full\_path1/file1.py\\
line: 10\\
class: MyClass1\\
line: 51\\
\verb|`|\verb|`|\verb|`|\\

Please think step by step before returning the location(s).\\
Your reasoning should start with "\#\#\# Thought:", and your answer should start with "\#\#\# Answer:".
        \end{exmp}
    \end{minipage}
\end{figure*}

\begin{figure*}[h]
    \begin{minipage}{\textwidth} 
        \begin{exmp}{Code Edit Generation Prompt}{code_edit}
You will be provided with an issue statement explaining a problem to resolve and a partial code base. Please first localize the bug based on the issue statement, and then generate \textit{*SEARCH/REPLACE*} edits to fix the issue.\\
\textit{\#\#\# GitHub Problem Description:}\\
\{issue\}

\textit{\#\#\# Code Content:}\\
\{code\_content\}

Please first localize the bug based on the issue statement, and then generate \textit{*SEARCH/REPLACE*} edits to fix the issue.\\

Every \textit{*SEARCH/REPLACE*} edit must use this format:\\
1. The file path\\
2. The start of search block: \textit{< < < < < < < SEARCH}\\
3. A contiguous chunk of lines to search for in the existing source code\\
4. The dividing line: =======\\
5. The lines to replace into the source code\\
6. The end of the replace block: \textit{> > > > > > > REPLACE}\\

For example:\\
\verb|`|\verb|`|\verb|`|python\\
\#\#\# mathweb/flask/app.py\\
\textit{< < < < < < < SEARCH}\\
from flask import Flask\\
=======\\
import math\\
from flask import Flask\\
\textit{> > > > > > > REPLACE}\\
\verb|`|\verb|`|\verb|`|

Please note that the \textit{*SEARCH/REPLACE*} edit REQUIRES PROPER INDENTATION. If you would like to add the line '        print(x)', you must fully write that out, with all those spaces before the code!
Wrap the \textit{*SEARCH/REPLACE*} edit in blocks \verb|`|\verb|`|\verb|`|python...\verb|`|\verb|`|\verb|`|.\\

Your reasoning should start with "\#\#\# Thought:", and your answer should start with "\#\#\# Answer:".
        \end{exmp}
    \end{minipage}
\end{figure*}

\end{document}